Destabilization of long-wavelength Love and Stoneley waves in slow sliding

K. Ranjith

Kulapura House, Pattanchery, Palakkad, Kerala 678532, India

E-mail: ranjith@post.harvard.edu

**Abstract**

Love waves are dispersive interfacial waves that are a mode of response for anti-plane motions of an elastic layer bonded to an elastic half-space. Similarly, Stoneley waves are interfacial waves in bonded contact of dissimilar elastic half-spaces, when the displacements are in the plane of the solids. It is shown that in slow sliding, long wavelength Love and Stoneley waves are destabilized by friction. Friction is assumed to have a positive instantaneous logarithmic dependence on slip rate and a logarithmic rate weakening behavior at steady-state.

Long wavelength instabilities occur generically in sliding with rate- and state-dependent friction, even when an interfacial wave does not exist. For slip at low rates, such instabilities are quasi-static in nature, i.e., the phase velocity is negligibly small in comparison to a shear wave speed. The existence of an interfacial wave in bonded contact permits an instability to propagate with a speed of the order of a shear wave speed even in slow sliding, indicating that the quasi-static approximation is not a valid one in such problems.

# 1. Introduction:

Destabilization of interfacial elastic waves due to friction has been a topic of some recent investigations (Adams, 1995, Ranjith and Rice, 2001). For in-plane elasticity problems, where displacements are confined to the plane of the solids, two well-known interfacial waves are the Stoneley wave (Stoneley, 1924) and the slip wave (Achenbach and Epstein, 1967). The Stoneley wave occurs in bonded contact of dissimilar elastic half-spaces while the slip wave, also called the generalized Rayleigh wave, is for a freely slipping interface between two half-spaces. There are no analogues of the Stoneley wave and the slip wave in anti-plane elasticity, where the displacement is normal to the plane of the solids. However, an interface wave solution does exist in the bonded contact of a *finite* layer on a half-space. This is the Love wave (Love, 1911). The Love wave differs from the Stoneley and slip waves in that (a) it always exists if the shear wave speed of the layer is greater than that of the half-space whereas the other two interfacial waves exist only when the shear wave speeds of the solids are not very different (b) its speed along the interface is greater than the shear wave speed of the layer but less than that of the substrate, while the other two waves are subsonic (c) it is dispersive and the dispersion relations are multi-valued.

In this paper, two problems are studied involving dissimilar materials that permit interfacial waves in bonded contact. Anti-plane sliding of a finite layer on an elastic half-space is first studied. In slow frictional sliding, it is shown that the Love wave is destabilized at long wavelengths. In-plane sliding of dissimilar elastic half-spaces is subsequently analyzed. It is shown that long wavelength Stoneley waves are also destabilized in slow sliding.

## 2. The anti-plane problem:

In this section, the equation governing the stability of steady sliding of an elastic layer on an elastic half-space are derived. The perturbations from steady sliding are assumed to be transverse to the direction of slip (i.e. anti-plane sliding). The elastodynamic relation between slip and shear stress perturbations is first derived. A friction law which also relates the slip and shear stress perturbations is then introduced. These two relations are used to obtained the equation governing slip stability.

Consider an isotropic elastic layer of thickness $h$ sliding on an isotropic elastic half-space with a steady rate $V_o$ (Figure 1). The steady motion is due to an applied shear stress $\tau_o$ which is at the friction threshold, $\tau_o = f\sigma_o$, where $\sigma_o$ is the compressive normal stress on the boundary of the layer and $f$ is the friction coefficient at slip rate $V_o$. The shear modulus, density and shear wave speed of the layer are denoted by $\mu$, $\rho$ and $c_s$, respectively, and corresponding properties of the half space are denoted by $\mu'$, $\rho'$ and $c'_s$.

A Cartesian coordinate system is located as shown in Figure 1 so that the interface between the solids is at $x_2 = 0$ and the layer slides in the $x_3$ direction. The elastic fields are assumed to be independent of the $x_3$ coordinate. We are interested in the relation between slip and stress perturbations at the interface when the perturbation is transverse to the direction of slip, namely in the $x_1$ direction. If $u_i(x_1, x_2, t)$, $i = 1, 2, 3$ denote the displacement field, due to isotropy of the solid, the only displacement component is that in the direction of slip:

$$u_1 = u_2 = 0$$
$$u_3 = u_3(x_1, x_2, t). \qquad (1)$$

Let $\tau_{ij}(x_1,x_2,t)$, $i,j = 1,2,3$ denote the stresses. The only non-zero stresses corresponding to the displacement field Eq. (1) are $\tau_{13} = \tau_{31}$ and $\tau_{23} = \tau_{32}$. They are given by

$$\tau_{13} = \mu \frac{\partial u_3}{\partial x_1}$$
$$\tau_{23} = \mu \frac{\partial u_3}{\partial x_2}, \qquad (2)$$

the latter being the traction component on planes normal to the $x_2$ direction.

The linear momentum balance equation for the layer is

$$\frac{\partial \tau_{13}}{\partial x_1} + \frac{\partial \tau_{23}}{\partial x_2} = \rho \frac{\partial^2 u_3}{\partial t^2}. \qquad (3)$$

Substituting for the stresses from Eq. (2), one gets the equation of motion

$$\frac{\partial^2 u_3}{\partial x_1^2} + \frac{\partial^2 u_3}{\partial x_2^2} = \frac{1}{c_s^2} \frac{\partial^2 u_3}{\partial t^2}, \qquad (4)$$

where $c_s = \sqrt{\mu/\rho}$. Similarly, the equation of motion of the elastic half-space in the region $x_2 < 0$ is

$$\frac{\partial^2 u_3}{\partial x_1^2} + \frac{\partial^2 u_3}{\partial x_2^2} = \frac{1}{c_s'^2} \frac{\partial^2 u_3}{\partial t^2}. \qquad (5)$$

where $c'_s = \sqrt{\mu'/\rho'}$ is the shear wave speed of the half space.

Slip at rate $V_o$ and a perturbation from it in a single Fourier mode of wavenumber $k$ can be represented by a displacement field of the form

$$u_3(x_1,x_2,t) = V_o t + U^+(k,p)e^{ikx_1}e^{\alpha x_2}e^{pt}, \quad x_2 > 0,$$
$$u_3(x_1,x_2,t) = U^-(k,p)e^{ikx_1}e^{\alpha' x_2}e^{pt}, \quad x_2 < 0. \tag{6}$$

where $p$ is a complex variable, dependent on $k$, which characterizes the time response to the perturbation. $\alpha(k,p)$ and $\alpha'(k,p)$ are to be determined so that the governing equations of motion are satisfied. Substituting into the equation of motion for the layer Eq. (4), gives

$$\alpha^2 = k^2 + \frac{p^2}{c_s^2}. \tag{7}$$

Defining

$$\alpha = |k|\sqrt{1 + p^2/k^2 c_s^2}, \tag{8}$$

where $\sqrt{\phantom{x}}$ denotes the analytic continuation of the positive square root function, both $\alpha$ and $-\alpha$ solve Eq. (7). A convenient choice of branch cuts in the complex $p$-plane is from the branch points $p = \pm i|k|c_s$ to $p = \infty$ along the imaginary axis, away from the origin. The general form of the displacement in the layer is therefore

$$u_3(x_1,x_2 > 0,t) = V_o t + [U_1^+(k,p)e^{-\alpha x_2} + U_2^+(k,p)e^{\alpha x_2}]e^{ikx_1}e^{pt} \tag{9}$$

The stress component $\tau_{23}$ in the layer corresponding to the above displacement field is

$$\tau_{23}(x_1,x_2 > 0,t) = \tau_o + \mu[-\alpha U_1^+(k,p)e^{-\alpha x_2} + \alpha U_2^+(k,p)e^{\alpha x_2}]e^{ikx_1}e^{pt} \tag{10}$$

The perturbations at the interface do not alter the applied shear stress $\tau_o$ on the boundary of the layer. Thus $\tau_{23}(x_1,h,t) = \tau_o$, so that

$$-U_1^+ e^{-\alpha h} + U_2^+ e^{\alpha h} = 0. \tag{11}$$

An analogous development for the half-space $x_2 < 0$ follows. The displacement field in the half-space is of the form

$$u_3(x_1, x_2 < 0, t) = U^-(k,p) e^{ikx_1} e^{\alpha' x_2} e^{pt}. \tag{12}$$

Substituting into the equation of motion for the half-space gives

$$\alpha'^2 = k^2 + \frac{p^2}{c_s'^2} \tag{13}$$

which has the solution

$$\alpha' = |k|\sqrt{1 + p^2/k^2 c_s'^2}. \tag{14}$$

Branch cuts are defined as before from $p = \pm i |k| c_s'$ to $p = \infty$ along the imaginary axis, away from the origin. This ensures that $\mathrm{Re}(\alpha') \geq 0$ for any $p$. It is noted that $-\alpha'$ is not a valid solution to Eq. (13) since it gives rise to an unbounded displacement field as $x_2 \to -\infty$.

The stress component $\tau_{23}$ in the half-space is then

$$\tau_{23}(x_1, x_2 < 0, t) = \tau_o + \mu' \alpha' U^-(k,p) e^{\alpha' x_2} e^{ikx_1} e^{pt} \tag{15}$$

The slip at the interface is

$$\begin{aligned}\delta(x_1, t) &= u_3(x_1, x_2 = 0^+, t) - u_3(x_1, x_2 = 0^-, t) \\ &= V_o t + [U_1^+ + U_2^+ - U^-] e^{ikx_1} e^{pt}. \end{aligned} \tag{16}$$

Denoting

$$D(k,p) = U_1^+(k,p) + U_2^+(k,p) - U^-(k,p), \tag{17}$$

the slip can be written as

$$\delta(x_1,t) = V_o t + D(k,p)e^{ikx_1}e^{pt}. \tag{18}$$

The traction component of stress at the interface

$$\tau(x_1,t) = \tau_{23}(x_1,0,t) \equiv \tau_o + T(k,p)e^{ikx_1}e^{pt} \tag{19}$$

is continuous. From Eq. (10) and Eq. (15) this requires

$$-\mu\alpha U_1^+ + \mu\alpha U_2^+ = \mu'\alpha' U^-. \tag{20}$$

Eqs. (11), (17) and (20) constitute a system of linear algebraic equations for $U_1^+$, $U_2^+$ and $U^-$ in terms of $D$. Solving that system,

$$U^- = -\frac{\mu\alpha}{\mu\alpha + \mu'\alpha'\coth\alpha h} D. \tag{21}$$

The shear stress at the interface is then

$$\tau(x_1,t) = \tau_o - \frac{\mu'\alpha'\mu\alpha}{\mu\alpha + \mu'\alpha'\coth\alpha h} D(k,p)e^{ikx_1}e^{pt}. \tag{22}$$

The amplitudes of the shear stress and slip perturbations at the interface thus satisfy

$$T(k,p) = -\frac{\mu'\alpha'\mu\alpha}{\mu\alpha + \mu'\alpha'\coth\alpha h} D(k,p). \tag{23}$$

When $h \to \infty$, corresponding to the anti-plane sliding of two dissimilar half-spaces,

$$T(k,p) = -\frac{\mu'\alpha'\mu\alpha}{\mu'\alpha' + \mu\alpha} D(k,p), \tag{24}$$

in agreement with the earlier result of Ranjith (2008) for that geometry. Writing

$$F(k,p) = \frac{2\mu'\sqrt{1 + p^2/k^2 c_s^2}\sqrt{1 + p^2/k^2 c_s'^2}}{\mu\sqrt{1 + p^2/k^2 c_s^2} + \mu'\sqrt{1 + p^2/k^2 c_s'^2}\coth(|k|h\sqrt{1 + p^2/k^2 c_s^2})}, \tag{25}$$

Eq. (23) takes the form

$$T(k,p) = -\frac{\mu|k|}{2}F(k,p)D(k,p). \tag{26}$$

For a given $k$, a pole of $F(k,p)$ indicates a stress perturbation with no associated slip perturbation. The only poles of $F(k,p)$ are zeroes of the function

$$M(k,p) = \mu\sqrt{1 + p^2/k^2 c_S^2} + \mu'\sqrt{1 + p^2/k^2 c_S'^2}\coth(|k|h\sqrt{1 + p^2/k^2 c_S^2}), \tag{27}$$

which is the equation for the Love wave in bonded contact of the layer and the half-space. Using the notation $c = -ip/k$ for the phase velocity, we focus on the properties of $F(k,c)$ and $M(k,c)$ when $c$ is real, corresponding to steady-state wave propagation. Also it is assumed without loss of generality that $c > 0$ - similar results apply for $c < 0$. The Love function $M(k,p)$ can then be written in terms of $c$ as

$$M(k,c) = i(\mu\sqrt{c^2/c_S^2 - 1} - \mu'\sqrt{1 - c^2/c_S'^2}\cot(|k|h\sqrt{c^2/c_S^2 - 1})). \tag{28}$$

It is readily seen that $M(k,c)$ has zeroes only when $c_S < c < c'_S$, corresponding to Love waves. The Love wave speed $c_o$ depends on the wavenumber $k$. The wave always exists for any $k$ and $\mu/\mu'$ as long as $c_S < c'_S$. In the long wavelength limit, $|k| \mapsto 0$, it is clear by inspection of Eq. (28) that $c_o \to c'_S$. In the short wavelength limit, $|k| \mapsto \infty$, there are multiple zeroes of the Love function due to the periodicity of the cotangent function. When $|k| \mapsto \infty$, the cotangent term in Eq. (28) has a limit only if $c \to c_S$. Since the first term in Eq. (28) also approaches zero as $c \to c_S$, the zeros, $c_n$, $n = 0,1,...,N(k)$, in the short wavelength limit occur close to the roots of the equation

$$|k|h\sqrt{c^2/c_S^2 - 1} = (2n+1)\pi/2, \quad |k| \mapsto \infty. \tag{29}$$

A zero of $F(k,c)$ indicates a slip perturbation with no associated stress perturbation. It is obvious that $F(k,c) = 0$ when $c = c_s$ and $c = c'_s$. However, these are branch points of $F(k,c)$, not zeroes, and represent 1-D body waves in either solid. For example, from Eq. (21) it is clear that when $c = c_s$, $U^- = 0$ and the displacement field is

$$u_3(x_1, x_2 > 0, t) = V_o t + D(k,p) e^{ik(x_1 + c_s t)},$$
$$u_3(x_1, x_2 < 0, t) = 0$$
(30)

The only zeroes of $F(k,c)$ are poles of the Love function $M(k,c)$. For generic $k$, poles occur only when $c_s < c < c'_s$ and they are determined by the condition that

$$|k| h \sqrt{c^2/c_s^2 - 1} = n\pi$$
(31)

for an integer $n \geq 1$. From Eq. (29) and Eq. (31), it is clear that the zeroes and poles of $F(k,c)$ alternate as $|k| \mapsto \infty$ with the first pole being closer to the branch point $c = c_s$ than the first zero.

Friction is now introduced at the interface and its effect on slip stability is studied. A friction law dependent on the slip rate $V(x_1, t)$ and a fading memory of its history, characterized by a state variable $\theta(x_1, t)$, is adopted. This is motivated by the experiments of Dieterich (1979) and Ruina (1983). At constant normal stress $\sigma_o$, the frictional shear stress is of the form

$$\tau = f(V, \theta) \sigma_o.$$
(32)

The above mentioned experiments involve application of step changes in slip velocity from steady sliding and observing the instantaneous as well as the gradual change of the frictional shear stress. A positive logarithmic instantaneous dependence of shear stress on the slip velocity was observed, i.e.,

$$\tau \sim a\ln(V)\sigma_o, \quad a > 0. \tag{33}$$

A logarithmic weakening with slip velocity at steady state was also seen in the experiments, so that at a steady slip velocity $V$, the frictional stress is

$$\tau = \tau_o - (b-a)\ln(V/V_o)\sigma_o, \quad b-a > 0. \tag{34}$$

It is observed that $b-a$ is of the same order as $a$. The gradual change of the frictional stress during an imposed step change in slip velocity, as seen in the experiments, is modeled empirically by the state variable $\theta(x_1,t)$. Ruina (1983) proposed a friction law of the form

$$\tau = \tau_o + a\ln(V/V_o)\sigma_o + b\theta\sigma_o,$$
$$\frac{\partial \theta}{\partial t} = -(V/L)(\theta + \ln(V/V_o))$$

for the change in shear stress from its value $\tau_o$ at the steady slip rate $V_o$. The constant $L$ is a characteristic length for evolution of the shear stress from $\tau_o$ to the steady value given by Eq. (34) in the velocity-stepping experiments. Linearizing the above friction law about the steady state and eliminating the state variable we obtain

$$\frac{\partial \tau}{\partial t} = \frac{a\sigma_o}{V_o}\frac{\partial V}{\partial t} - \frac{V_o}{L}\left(\tau - \tau_o + \frac{(b-a)\sigma_o}{V_o}(V - V_o)\right). \tag{35}$$

Noting that $\tau - \tau_o = T(k,p)\exp(ikx_1 + pt)$ and $V - V_o = pD(k,p)\exp(ikx_1 + pt)$, the linearized friction law Eq. (35) reduces to

$$\left(p + \frac{V_o}{L}\right)T(k,p) = \frac{\sigma_o}{V_o}\left(ap - (b-a)\frac{V_o}{L}\right)pD(k,p). \tag{36}$$

Using the elastic relation between $T(k,p)$ and $D(k,p)$, Eq. (26), in the above, we get the equation for stability as

$$\frac{\mu|k|}{2}\left(p+\frac{V_o}{L}\right)F(k,p)+\frac{\sigma_o}{V_o}\left(ap-(b-a)\frac{V_o}{L}\right)p=0. \qquad (37)$$

For a given wavenumber $k$, a root of the above equation at $p = p_1 + ip_2$ indicates a slip response of the form

$$\delta(x_1,t)-V_o t \sim e^{ikx_1}e^{p_1 t}e^{ip_2 t} = e^{ik(x_1+p_2 t/k)}e^{p_1 t}.$$

Thus, a root with a positive real part, $p_1 > 0$, indicates unstable slip. The phase velocity is clearly $c = -p_2/k$. We say that the slip response is quasi-static if the phase velocity magnitude is negligibly small in comparison to a shear wave speed, $|c|/c_s \ll 1$.

The following non-dimensional parameters and variables

$$\begin{aligned} K &= \frac{\mu|k|L}{2a\sigma_o}, \\ S &= p/|k|c_s, \\ H &= \frac{2a\sigma_o h}{\mu L}, \\ \varepsilon &= \frac{\mu V_o}{2a\sigma_o c_s}, \end{aligned} \qquad (38)$$

are now introduced. $K$ is a non-dimensional wavenumber and $H$, a non-dimensional layer thickness. The non-dimensional $S$ used above is particularly convenient since its imaginary part gives the phase velocity in comparison to the shear wave speed of the layer. The non-dimensional slip velocity $\varepsilon$ can be thought of as a measure of the elastodynamic stress change in relation to the frictional stress change accompanying a small slip velocity change $\Delta V$ from steady sliding at rate $V_o$. The former is $(\mu/2c_s)\Delta V$ while the latter is $(a\sigma_o/V_o)\Delta V$. When $\varepsilon \ll 1$, i.e. slip velocity is sufficiently low, the elastodynamic stress change is small and it may be naively expected that elastodynamic effects would be negligible. However, as shown in the following, that is not generally the case.

Using the non-dimensional quantities in Eq. (38), the governing equation for stability can be written as

$$\left(1+\frac{SK}{\varepsilon}\right)\frac{2\mu'\sqrt{1+S^2}\sqrt{1+S^2\frac{c_S^2}{c_S'^2}}}{\mu\sqrt{1+S^2}+\mu'\sqrt{1+S^2\frac{c_S^2}{c_S'^2}}\coth\left(KH\sqrt{1+S^2}\right)}+\frac{S}{\varepsilon}\left(\frac{SK}{\varepsilon}-\frac{b-a}{a}\right)=0 \qquad (39)$$

or

$$\left(1+\frac{SK}{\varepsilon}\right)F(K,S)+\frac{S}{\varepsilon}\left(\frac{SK}{\varepsilon}-\frac{b-a}{a}\right)=0. \qquad (40)$$

### 3. Stability analysis:

In this section, slip stability in slow anti-plane sliding, $\varepsilon \ll 1$, is investigated for short and long wavelength perturbations. It is shown that the response to short wavelength perturbations is stable, thus ensuring that the stability problem is well-posed. The response to long wavelength perturbations is however shown to be generically unstable. In particular, long wavelength Love waves are shown to be destabilized in slow sliding.

The short wavelength limit is given by $\varepsilon \ll 1 \ll K$ while the long wavelength limit is $K \ll \varepsilon \ll 1$. First, short wavelength stability is studied. Eq. (40) can be written as

$$\left[\frac{SK}{\varepsilon}\left(F(K,S)+\frac{S}{\varepsilon}\right)\right]+\left[F(K,S)-\frac{S}{\varepsilon}\frac{(b-a)}{a}\right]=0 \tag{41}$$

and we look for solutions $S$ that are successively $O(\varepsilon/K)$, $O(\varepsilon)$, and $O(1)$. When $S=O(\varepsilon/K)$, $F(K,S)=O(1)$ and it is easily verified that there are no solutions of that order.

When $S=O(\varepsilon)$, $F(S,K)=O(1)$ again and the balance of terms becomes

$$F(K,S)+\frac{S}{\varepsilon}=0. \tag{42}$$

The root of the above equation is

$$\begin{aligned}S&=-\varepsilon F(K\to\infty,0)\\&=-2\varepsilon\frac{\mu'}{\mu+\mu'}\end{aligned} \tag{43}$$

Clearly $\mathrm{Re}(S)<0$ and the root is stable. Next we look for roots $S=O(1)$. The balance of terms in Eq. (41) again leads to Eq. (42). As discussed in the previous section, when $S=O(1)$, $F(K,S)$ has poles that correspond to Love waves. It was seen that in the large $K$ limit, $F(K,S)$ has multiple poles along the imaginary $S$-axis, $S=\pm iC_n=\pm ic_n/c_S$, $n=0,1,2,...,N(K)$, close to the roots of

$$KH\sqrt{C^2-1}=(2n+1)\pi/2,\quad K\to\infty. \tag{44}$$

Close to $S=iC_n$, $F(K,S)$ has the structure

$$F(K,S)=\frac{iA_n}{S-iC_n} \tag{45}$$

where, by inspection of $F(K,S)$, $A_n=A_n(K,C_n)$ is a real constant. (For the pole at $S=-iC_n$, the sign of $A_n$ changes). Therefore, Eq. (42) is of the form

$$\frac{iA_n}{S-iC_n}+\frac{S}{\varepsilon}=0. \tag{46}$$

The roots are therefore

$$S=\pm iC_n - \frac{A_n}{C_n}\varepsilon. \tag{47}$$

To ensure stability at short wavelengths, we need to show that each $A_n > 0$. It was noted earlier that the poles and zeroes of $F(K \to \infty, S)$ alternate. Therefore, $A_n$ is of the same sign for every $S = iC_n$ corresponding to a given, large $K$ and it suffices to show that the coefficient corresponding to the fundamental mode $A_o > 0$. Observing that very close to $S = \pm i$, the cotangent term in Eq. (28) dominates, the singular structure has to be such that $A_o > 0$. Thus stability of short wavelength perturbations is ensured.

In the long wavelength limit, $K \ll \varepsilon \ll 1$, so that $\varepsilon \ll 1 \ll \varepsilon/K$. We look for solutions $S$ that are $O(\varepsilon)$, $O(1)$ and $O(\varepsilon/K)$ as before. At $O(\varepsilon)$, the dominant terms give

$$F(K,S) - \frac{(b-a)}{a}\frac{S}{\varepsilon} = 0 \tag{48}$$

which has the solution

$$S = \frac{a}{b-a}\varepsilon F(K \to 0, 0). \tag{49}$$

But $F(K \to 0, 0) = O(K)$ and therefore there are no roots that are $O(\varepsilon)$. When $S = O(1)$, we again get Eq. (48). It has been pointed out that when $K \to 0$, $F(K,S)$ has a pole at $S = \pm iC_o = \pm ic_o/c_s$, with $c_o$ being close to $c'_s$, corresponding to the Love wave. As discussed earlier, the singular structure close to the pole is

$$F(K,S) = \frac{iA_o}{S - iC_o}. \tag{50}$$

The roots of Eq. (48) are therefore

$$S = \pm iC_o + \frac{aA_o}{(b-a)C_o}\varepsilon \tag{51}$$

The argument previously made for $A_o$ being positive still holds and therefore long wavelength perturbations are unstable with velocity weakening friction, $b - a > 0$. The speed of propagation of the wave is precisely that of the Love wave.

When $S = O(\varepsilon/K)$, the balance of terms gives

$$\frac{SK}{\varepsilon} - \frac{b-a}{a} = 0 \tag{52}$$

so that the root is

$$S = \frac{b-a}{a}\frac{\varepsilon}{K}, \tag{53}$$

indicating instability. However, the phase velocity is zero to leading order, indicating the quasi-static nature of the instability. It must be noted that unstable roots at $O(\varepsilon)$ and $O(\varepsilon/K)$ generically occur in frictional stability problems. For the simple case of anti-plane sliding of identical elastic half-spaces, Eq. (40) becomes

$$\left(1 + \frac{SK}{\varepsilon}\right)\sqrt{1+S^2} + \frac{S}{\varepsilon}\left(\frac{SK}{\varepsilon} - \frac{b-a}{a}\right) = 0. \tag{54}$$

When $K \ll \varepsilon \ll 1$, this equation has the solutions

$$\begin{aligned} S &= \frac{a}{b-a}\varepsilon, \\ S &= \frac{b-a}{a}\frac{\varepsilon}{K}. \end{aligned} \tag{55}$$

## 4. The in-plane problem:

In this section, the stability of slow sliding of dissimilar elastic half-spaces is studied when the perturbations are in the direction of slip (i.e. in-plane sliding). The elastodynamic relations for this problem have been derived by Ranjith and Rice (2001) and slow slip stability for short-wavelength perturbations has been studied by Rice et al. (2001). Here, attention is focused on the long-wavelength limit. It is shown that long-wavelength Stoneley waves are destabilized in slow sliding.

As shown in Figure 2, a Cartesian coordinate system is located so that the interface is at $x_2 = 0$ and steady sliding with rate $V_o$ occurs in the $x_1$ direction. The elastic fields are assumed to be independent of the $x_3$ coordinate. The far-field applied stresses are $\tau_{21} = \tau_o$ and $\tau_{22} = -\sigma_o$ such that they are at the friction threshold, $\tau_o = f\sigma_o$. At steady state, the shear and normal stresses at the interface equal the far-field values.

Interfacial slip representing steady sliding with rate $V_o$ and a perturbation from it in a single Fourier mode of wavenumber $k$ is of the form

$$\delta(x_1,t) = V_o t + D(k,p)e^{ikx_1}e^{pt}. \tag{56}$$

The corresponding elastic shear and compressive normal stresses on the interface are given by

$$\begin{aligned}\tau &= \tau_o + T(k,p)e^{ikx_1}e^{pt} \\ \sigma &= \sigma_o - \Sigma(k,p)e^{ikx_1}e^{pt}\end{aligned} \tag{57}$$

where

$$T(k,p) = -\frac{\mu|k|}{2}Y_{11}(k,p)D(k,p),$$

$$\Sigma(k,p) = -\frac{\mu|k|}{2}Y_{21}(k,p)D(k,p). \tag{58}$$

The explicit forms of $Y_{11}(k,p)$ and $Y_{21}(k,p)$ in terms of the elastic properties and wave speeds of the solids are given in Ranjith and Rice (2001). It is noted that due to the difference in material properties across the interface, the slip perturbation induces a normal stress change at the interface in addition to a shear stress change.

Since slip couples with normal stress, a friction law including the dynamic response to normal stress changes is needed. Rice et al. (2001) proposed a general linear friction law of the form

$$\frac{\partial \tau}{\partial t} = (f-\alpha)\frac{\partial \sigma}{\partial t} + \frac{a\sigma_o}{V_o}\frac{\partial V}{\partial t} - \frac{V_o}{L}\left(\tau - \tau_o - f(\sigma - \sigma_o) + \frac{(b-a)\sigma_o}{V_o}(V-V_o)\right). \tag{59}$$

Here $f$ and $\alpha$ are constants. (The $\alpha$ above is not to be confused with the $\alpha$ defined in Eq. (8), which is not used in the following). The first term on the right hand side above is the Coulomb-type instantaneous response to a normal stress change and the term in the square brackets incorporates a memory of normal stress history. Using this friction law, Rice et al. (2001) showed that the equation governing slip stability is

$$\left(1+\frac{SK}{\varepsilon}\right)Y_{11}(S) + \left(f+(f-\alpha)\frac{SK}{\varepsilon}\right)Y_{21}(S) + \frac{S}{\varepsilon}\left(\frac{SK}{\varepsilon} - \frac{b-a}{a}\right) = 0. \tag{60}$$

The function $Y_{11}(S)$ has zeros corresponding to the slip wave, when it exists, and both $Y_{11}(S)$ and $Y_{21}(S)$ have poles, corresponding to the Stoneley wave, when it exists.

For slow sliding, the short wavelength limit of Eq. (60) was studied in Rice et al. (2001). When $\varepsilon \ll 1 \ll K$, the roots of Eq. (60) were shown to occur at $O(\varepsilon)$ and $O(1)$. At $O(\varepsilon)$, the balance of terms gives

$$Y_{11}(S) + (f - \alpha)Y_{21}(S) + \frac{S}{\varepsilon} = 0 \tag{61}$$

which has the root

$$S = -\varepsilon\big(Y_{11}(0) + (f - \alpha)Y_{21}(0)\big). \tag{62}$$

Rice et al. (2001) showed that the real part of this root is negative. Hence it is stable. The roots at $O(1)$ occur close to the Stoneley poles. The dominant terms are again those in Eq. (61). The singular structure close to the Stoneley pole at $S = iC_{St}$ can be written as

$$\begin{aligned} Y_{11}(S) &= \frac{iA}{S - iC_{St}}, \\ Y_{21}(S) &= \frac{B}{S - iC_{St}}. \end{aligned} \tag{63}$$

where $A$ and $B$ are real constants. (For the Stoneley pole at $S = -iC_{St}$, the sign of $A$ and $B$ changes). The roots close to the Stoneley poles are then

$$S = \pm iC_{St} - \varepsilon\left(\frac{A}{C_{St}} - i\frac{(f - \alpha)B}{C_{St}}\right). \tag{64}$$

Rice et al. (2001) showed, using general arguments, that whenever the Stoneley pole exists, $A > 0$. Therefore the roots at $O(1)$ are also stable.

The long wavelength limit of Eq. (60), $K \ll \varepsilon \ll 1$, is now studied. We look for solutions $S$ that are $O(\varepsilon)$, $O(1)$ and $O(\varepsilon/K)$. At $O(\varepsilon)$, the governing equation is

$$Y_{11}(S) + fY_{21}(S) - \frac{S}{\varepsilon}\frac{b-a}{a} = 0 \tag{65}$$

which has the solution

$$S = \varepsilon\frac{b-a}{a}\bigl(Y_{11}(0) + fY_{21}(0)\bigr). \tag{66}$$

Comparing to Eq. (62), which was shown to have a negative real part, we conclude that the above root has a positive real part for velocity weakening friction, $b-a>0$. Hence the root is unstable. At $O(1)$, the governing equation remains Eq. (65) and we expect roots close to the Stoneley poles. Assuming the singular structure as in Eq. (63), the roots can be shown to be

$$S = \pm iC_{St} + \varepsilon\frac{a}{b-a}\left(\frac{A}{C_{St}} - i\frac{(f-\alpha)B}{C_{St}}\right). \tag{67}$$

As mentioned, Rice et al. (2001) showed that $A>0$. Hence the Stoneley wave is destabilized at long wavelengths. Finally at $O(\varepsilon/K)$, the root can be shown to be

$$S = \frac{b-a}{a}\frac{\varepsilon}{K} \tag{68}$$

which is also unstable when $b-a>0$. In summary, it has been shown that slip response at long wavelengths is always unstable. The roots given by Eq. (66) and Eq. (68) have zero imaginary parts (to their respective orders). When a Stoneley wave exists, it is destabilized at long wavelengths.

## 5. Discussion:

Solid mechanics problems involving slowly moving boundaries are often studied using the quasi-static approximation, i.e. stress transfers are assumed to be instantaneous and not wave-mediated. However, this approximation is not always valid. An example is the slow growth of cracks, as in fatigue, which is generally assumed to be a quasi-static process. A recently discovered instability of crack fronts (Ramanathan and Fisher, 1997, Morrissey and Rice, 1998) shows that elastodynamic effects can be important even in slow crack growth. The present study has an analogous implication for sliding, i.e., slow sliding is not in general the same as quasi-static sliding.

It is instructive to explicitly analyze the stability of slow sliding in the quasi-static approximation. The governing equation for quasi-static anti-plane deformation is the Laplace equation,

$$\frac{\partial^2 u_3}{\partial x_1^2} + \frac{\partial^2 u_3}{\partial x_2^2} = 0. \qquad (69)$$

Consider the problem geometry as in Figure 1. For anti-plane sliding of an elastic layer on a dissimilar elastic half-space, the quasi-static elastic relation between shear stress and slip perturbations from steady state, analogous to Eq. (23) for the dynamic case, can be shown to be

$$T(k,p) = -\frac{\mu |k|}{2} \frac{2\mu'}{\mu + \mu' \coth(|k|h)} D(k,p). \qquad (70)$$

The governing equation for slip stability becomes, following the steps leading to Eq. (37),

$$\frac{\mu |k|}{2}\left(p+\frac{V_o}{L}\right)\left(\frac{2\mu'}{\mu+\mu'\coth(|k|h)}\right)+\frac{\sigma_o}{V_o}\left(ap-(b-a)\frac{V_o}{L}\right)p=0 \qquad (71)$$

In the long wavelength limit, $|k| \mapsto 0$, this yields a quadratic equation for $p$ in terms of $|k|$ as

$$ap^2+\left(\frac{\mu V_o |k|^2 h}{\sigma_o}-\frac{(b-a)V_o}{L}\right)p+\frac{\mu V_o^2 |k|^2 h}{L\sigma_o}=0 \qquad (72)$$

For sufficiently long wavelengths, it is clear that the roots of the above equation are

$$p=\frac{(b-a)}{a}\frac{V_o}{L}+O(|k|^2) \text{ and}$$
$$p=O(|k|^2), \qquad (73)$$

both being real and positive. Thus the response at long wavelengths is unstable with the phase velocity of the instability being zero to leading order. It has been shown in this paper that the above quasi-static behavior does *not* emerge as a limit of the full elastodynamic equations when the slip velocity is low. An additional unstable root occurs when elastodynamic effects, representing wave-mediated stress transfers, are included. From Eq. (51), this root is

$$p=\frac{A_o \mu c_s V_o}{2(b-a)\sigma_o c_o}|k|\pm ic_o|k|+O(|k|^2) \qquad (74)$$

where $c_o$ is the Love wave speed in the long wavelength limit (approximately $c_s'$, the shear wave speed of the substrate) and $A_o$ is a positive constant. A similar discussion applies for the in-plane sliding problem of Section 4, with the Stoneley wave playing a role analogous to that of the Love wave above. It must be noted that in the sliding of *identical half-spaces*, a quasi-static limit for slow sliding does exist, as shown in Rice et al. (2001). In that case, the existence of Love and Stoneley waves is precluded and the results of the present analysis do not carry over.

Physically, the results obtained here have the surprising implication that even surfaces that are slowly slid can produce acoustic emissions. Prior work, summarized by Rice et al. (2001), had suggested that such would not be the case. As mentioned, those studies assumed geometries and material properties which precluded the existence of interfacial waves in bonded contact. In the context of the earth, the instability identified here could be a possible origin of the observed global seismic background radiation. It is well known that large earthquakes excite the free oscillations of the earth. However, Nawa et al. (1998) have reported that the fundamental long-period spheroidal modes of the earth's oscillations are continuously excited, even when large earthquakes do not occur. Similar continuous excitation of the fundamental toroidal modes has recently been observed by Kurrle and Widmer-Schnidrig (2008). The present analysis suggests that destabilization of Love and Stoneley waves in the slow sliding of tectonic plates, away from large earthquakes, could be a possible mechanism for the continuous excitation of the earth's oscillations. The spheroidal modes involve vertical motions of the surface, consistent with the displacements caused by Stoneley waves, and horizontal motions accompany the toroidal modes as is characteristic of Love waves.

## 6. Conclusions:

It has been shown that long wavelength Love and Stoneley waves are destabilized in slow frictional sliding. Essential to the analysis is the assumption that friction has a logarithmic dependence on slip rate, both instantaneously and in the steady state, the former being positive and the latter being negative, but both effects being of the same order. Thus, the quasi-static approximation is not a valid one for slow sliding if the geometry of the problem and material properties are such that an interfacial wave exists in bonded contact of the solids.

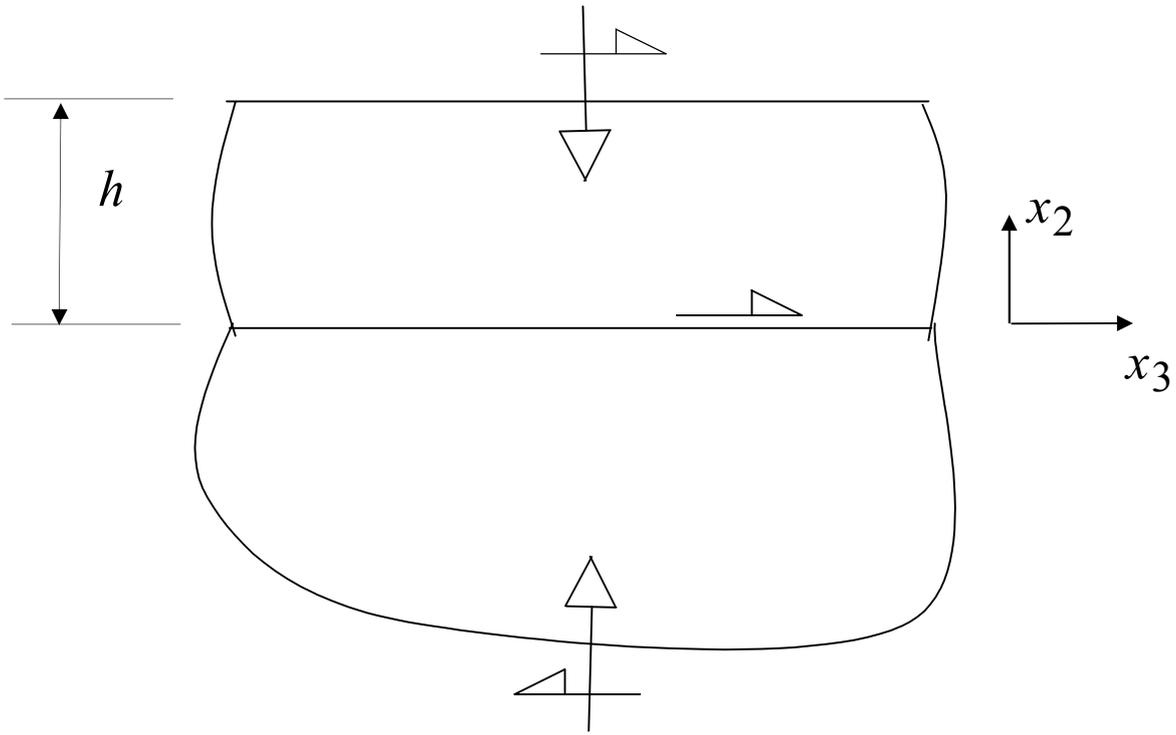

Figure 1: Geometry for the anti-plane sliding problem

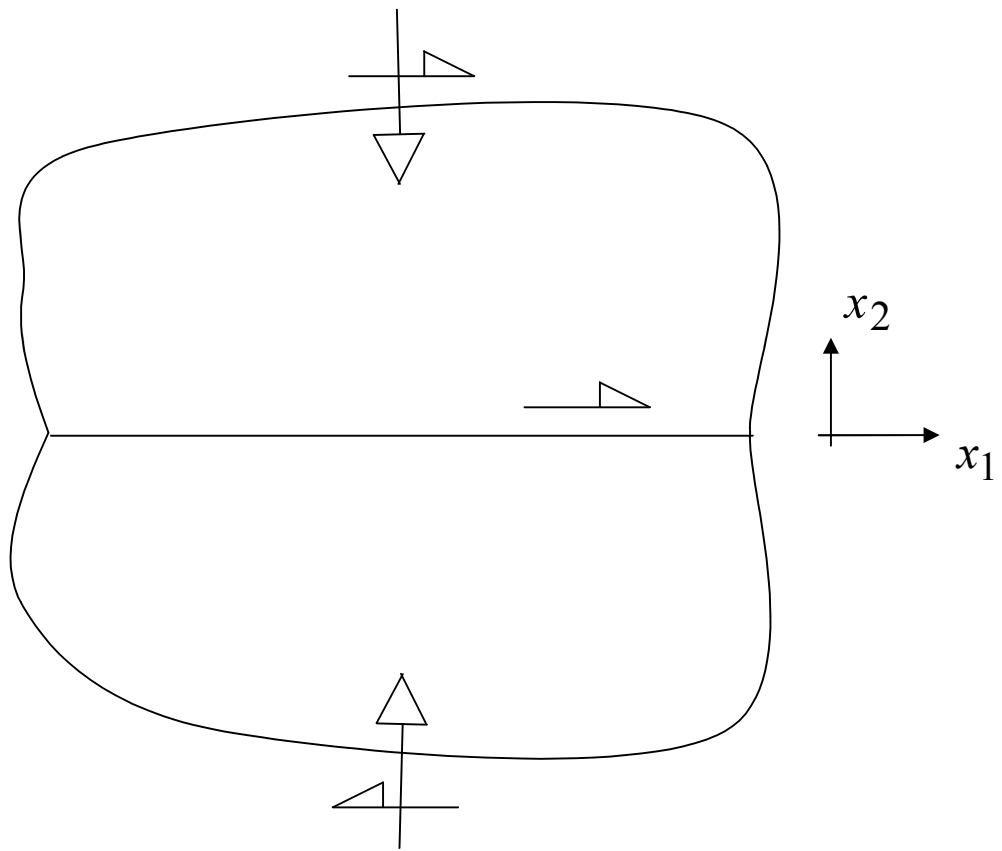

Figure 2: Geometry for the in-plane sliding problem